\pdfoutput=1		
\documentclass[12pt,a4paper,twoside,onecolumn]{article}

\usepackage{graphicx}           	
\usepackage{caption}            	
\usepackage{fancyhdr}           	
\usepackage{latexsym}           	
\usepackage{amssymb}            	
\usepackage{amsbsy}		
\usepackage{endnotes}          

\usepackage{fourier}            
 
\newcommand{\ignore}[1]{}   
\newcommand{\0}[1]{\phantom{0}}    

\begin{document}
\small       
\sloppy      
\widowpenalty=300
\clubpenalty=300

\parindent 0truecm
\null
\vfill

\hrule height 2pt
\vspace{10pt}
{\huge The Origin of the Solar System}

\vspace{15pt}

\hrule

\vspace{20pt}
{\large MICHAEL PERRYMAN}

\vspace{10pt}
School of Physics, University of Bristol

E-mail: mac.perryman@gmail.com

\vspace{20pt}
This article relates two topics of central importance in modern astronomy -- the discovery some fifteen years ago of the first planets around other stars (exoplanets), and the centuries-old problem of understanding the origin of our own solar system, with its planets, planetary satellites, asteroids, and comets. The surprising diversity of exoplanets, of which more than 500 have already been discovered, has required new models to explain their formation and evolution. In turn, these models explain, rather naturally, a number of important features of our own solar system, amongst them the masses and orbits of the terrestrial and gas giant planets, the presence and distribution of asteroids and comets, the origin and impact cratering of the Moon, and the existence of water on Earth.

\parindent 0.5truecm
\section*{Introduction}

After centuries of philosophical speculation about the existence of worlds beyond our solar system, the first hints of objects of planetary mass orbiting other stars were reported in the late 1980s. In 1992, a planetary system was discovered around a `millisecond pulsar' -- a rapidly spinning neutron star in the terminal, burnt out stages of stellar evolution. Then, in 1995, from repeated and very accurate telescope measurements of the motion of the 5th magnitude star 51~Peg in the constellation of Pegasus, convincing evidence for the first planet surrounding a more normal star was announced.

Two further exoplanets were known at the end of that year, and 34 at the end of the millennium. Since then, in just ten years, around 500 have been discovered through various methods, the 500th announced in December 2010. A remarkable advance in the understanding of their physical, chemical and dynamical properties, and their formation and evolution, has kept pace with this accelerating discovery. 

These new theoretical refinements have, in turn, confirmed the basic understanding of the formation of our own solar system, but they have also thrown important new light on how certain of its fundamental features have come about\endnote{Further details of all of the topics covered in this article, and comprehensive references to the work cited, can be found in {\it The Exoplanet Handbook} by Michael Perryman, Cambridge University Press (2011). An on-line compilation of the current status of exoplanet discoveries, The Extrasolar Planets Encyclopedia, is given at http://exoplanet.eu.}.

\vspace{10pt}
\hrule
\vspace{10pt}
\noindent
Invited contribution to the {\it European Review}, the Interdisciplinary Journal of the Humanities and Sciences of the Academia Europ{\ae}a.
Submitted May 2011. 

\clearpage
\section*{Detecting other planets}
The direct imaging of exoplanets, in the sense of observing a distant planet as a point of light reflected from its host stars, is an enormously difficult observational challenge for astronomers. An Earth-like (or even a Jupiter-like) planet orbiting another star would be millions of times fainter than the star itself, and the planet would orbit so close (in angular terms) such as to be completely `buried' within the stellar glare. A loose analogy would be trying to detect a moth dancing around a lighthouse seen from several kilometers distance. In the last few years, a handful of exoplanets have actually been detected as point sources of reflected light, but they are massive planets, orbiting very far out in their respective solar systems. The direct imaging of an Earth-mass exoplanet in an Earth-like orbit will be extremely difficult, and may be not be achieved for many years to come.

\begin{figure}[t]
\centering
\includegraphics[width=0.55\linewidth]{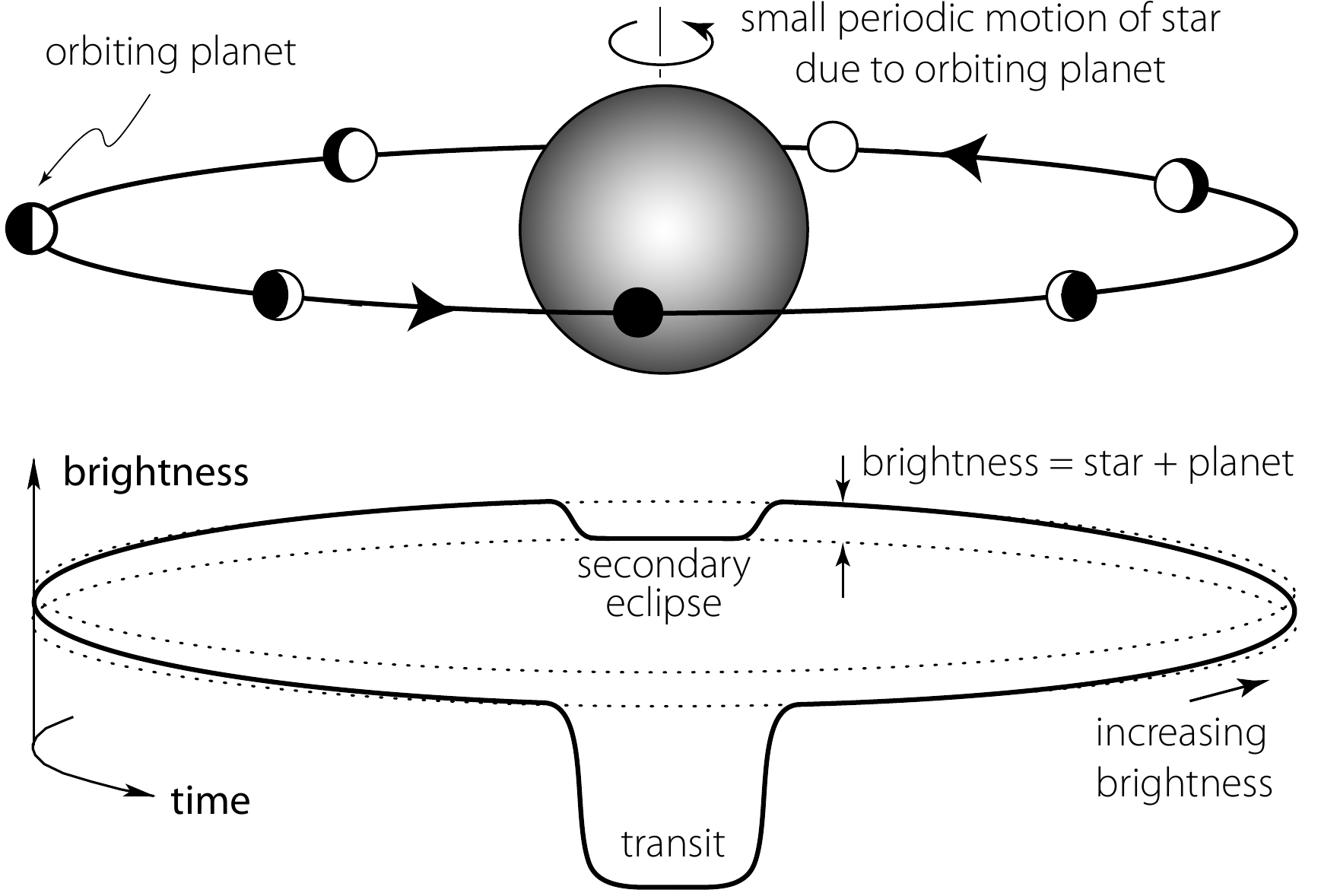} \hfill
\includegraphics[width=0.37\linewidth]{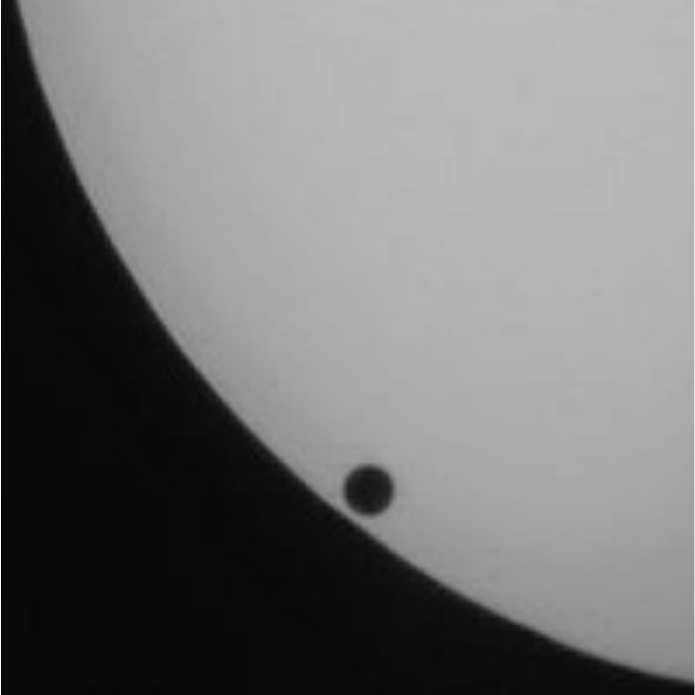}
\caption{\footnotesize Left upper: a planet orbiting another star causes the star to oscillate, slightly but periodically, due to the gravitational force between the star and planet. The star moves backwards and forwards along the line-of-sight as the planet orbits, allowing the planet's presence to be inferred from tiny periodic changes in the wavelength of the star's spectral lines -- this is the `radial velocity' or `Doppler' detection method. 
Left lower: if the line-of-sight from the Earth to the star happens to intersect the plane of the planet's orbit, the planet may be observed to transit in front of the star, thus blocking a fraction of the star light. After the transit, the planet's brighter day-side progressively comes into view, and the total brightness rises very slightly. The total light drops again during the secondary eclipse as the planet passes behind the star. Right: the transit of Venus, as its starts to cross the face of the Sun, observed in 2004. For exoplanet transits, the planet cannot be resolved, but the dimming of the star light can still be measured.
\label{fig:schematic-rv-transit}
}
\end{figure}

However, through various more subtle observational techniques, using astronomical telescopes both on the ground and in space, other effects of a planet's presence can be discerned. In one of the most important of these techniques (Figure~\ref{fig:schematic-rv-transit}, upper), the orbiting planet is sensed through the gravitational forces acting on its host star. This is possible because, strictly, a planet does not orbit the fixed star but, rather, both planet and star orbit their common centre of mass (somewhat like a pair of ice skaters circling each other whilst holding hands). The star's much larger mass means that it undergoes a much smaller wobble than the planet, but with the same orbital period. By carefully observing the wavelength of spectral lines emitted by the star over months or years, the star's radial motion (or Doppler shift) can reveal the presence of an orbiting planet. As one cannot know, in advance, which stars in the sky might harbour planets, searches must be conducted around many stars, and for the many months or years required for a repeating orbital patterns to be revealed. More than two hundred planets have now been found by this technique, and as many as 10\% or more of the stars visible to the naked eye are now known to have planets in orbit around them. 

The mass of the planet can be estimated from the velocity of the star's motion. Fifteen years ago, state-of-the-art spectroscopic measurement accuracies permitted the discovery of the first exoplanets with masses some several times that of Jupiter\endnote{Jupiter is the most massive of the solar system planets, some 300 times the mass of the Earth. It orbits the Sun, at a distance just over five times that of Earth, in about 12~years. The other `gas giant' planets (Saturn, Uranus, and Neptune) range between 15--35~Earth masses, with Neptune orbiting the Sun at some 30~times the distance of the Earth in about 160~years. The other `terrestrial' planets (Mercury, Venus, and Mars), are all less massive than Earth, with Mercury being the lightest (at one sixteenth that of Earth) and orbiting closest to the Sun in just 88~days.}. Today, as a result of a concerted effort of some 30 astronomical instrumentation groups worldwide, but most successfully by the Swiss-led HARPS spectrograph at the European Southern Observatory's 3.6-m telescope in Chile, and the US HIRES instrument at the Keck 10-m telescope in Hawaii, planets with masses closely resembling that of the Earth are now being found. A planet with the mass of the Earth can only be detected if the star's motion can be measured with an accuracy better than a meter per second, a remarkable achievement corresponding to a leisurely walking pace even over the gulfs of space separating our solar system from the target stars.

In another important technique, the brightness of a chosen target star is carefully monitored. If a planet is present, and if the plane of its orbit happens to lie along the line-of-sight to the star, its presence may be inferred from the tiny decrement in star light when the planet chances to pass across the face of the star (Figure~\ref{fig:schematic-rv-transit}, lower). The effect is exactly the same as that seen in the transits of Mercury or Venus as seen from the Earth\endnote{As seen from Earth, Mercury transits the face of our Sun 13 or 14 times per century, the last time in 2006, and the next on 9~May 2016. Transits of Venus are rarer, with pairs eight years apart separated by intervals of more than 100~years; the last occurred on 8~June 2004, the next will occur on 6~June 2012, but thereafter only in December 2117. These somewhat complex transit event patterns result from the fact that we are viewing one transiting planet from another orbiting planet.}, although only the amount of dimming can be measured, while the planet's passage across the star's disk cannot itself be resolved. A typical transit signature might last for several hours. But the star must be monitored for many weeks, or even many years, for the repeating pattern signifying the orbiting planet to be securely established. This particular approach of searching for `photometric transits' has revealed more than 100 planets since the first transiting exoplanet was discovered in 1999. Amongst the most prolific discoveries are those of the UK-led WASP consortium and the Hungarian-led HAT consortium using ground-based observatories and, from space, the French-led CoRoT satellite launched in 2006 and the US Kepler satellite launched in 2009.

Over the past decade, the transit method has developed into an immensely important diagnostic in the tool-kit of exoplanet hunters. This is because, while the mass of the planet can be measured from its gravitational tug on its parent star based on the Doppler measurements, its diameter can be deduced from the fractional amount of dimming as it transits -- the bigger the planet, the more star light is blocked, and the greater is the degree of dimming. These two basic planetary properties -- its mass, and its size -- allow a robust estimate of its density and therefore an extremely important clue as to its internal composition. More delicately, and indeed more remarkably, observing the changes in the spectrum of the star as the planet passes across its face (and, half an orbital period later, as the planet passes behind the star), the major chemical constituents of its atmosphere can be measured. In this way, astronomers are starting to get the first clues as to the composition of exoplanet atmospheres.

Various other exoplanet detection techniques are being developed and applied. In one of the most curious, planets can be revealed through the chance and exceedingly rare alignment between a distant planetary system and an even more distant background star. In a remarkable application of Einstein's general theory of relativity, which Einstein himself considered but quickly dismissed as unfeasible in practice, a planetary signature appears fleetingly through the process of gravitational `lensing' when the alignment is almost perfect. The first planet such was discovered in 2004 and, amongst tens of millions of stars now being scrutinised, a total of twelve have now been discovered.

\section*{The formation of planetary systems}
Historically, it has been a remarkably difficult challenge to come up with a satisfactory and compelling model of the formation of our own solar system. Indeed, as the astonishing wealth of detailed information about it has advanced in recent years, from a huge variety of ground-based observations, as well as a progression of remarkable fly-by, orbiting, and lander satellite missions, the observational data has become ever more complex, and the theoretician's task has become ever harder. 

Scientific theories of the solar system formation date back to the works of Emanuel Swedenborg in 1734, the Compte de Buffon in 1749, Immanuel Kant in 1755, and Pierre-Simon Laplace in 1796. Laplace considered that gaseous clouds, or nebulae, slowly rotate, gradually collapse and flatten due to gravity, and eventually form stars and planets. In this model, planets formed by detachment of a discrete system of gaseous rings, within which clumps and planets grew by spontaneous gravitational collapse.  Laplace's nebula model was pre-eminent in the 19th century. But it was largely abandoned in the early 20th century because it was unclear how material could be spontaneously partitioned such that the Sun retained 99.86\% of the mass of the solar system while the planets ended up with 99.5\% of the total angular momentum in their orbital motion.

Various alternatives were subsequently proposed. These included the Chamberlin--Moulton theory (1905) in which embryonic `planetesimals' formed out of material from an erupting Sun, a model involving tidal interactions between the Sun and a massive star (Jeans, 1917), the star--Sun collision theory (Jeffreys, 1929), the Schmidt--Lyttleton cloud accretion model, the turbulence-collision driven protoplanet theory due to McCrea, and Woolfson's tidal capture theory involving the Sun and a cool low-mass protostar. 

Today, the most widely-accepted model of the formation of our own solar system, which extends to the basic paradigm of exoplanet formation more generally, is referred to as the `solar nebula hypothesis'. According to this understanding, planets formed within the disk of gas (mainly hydrogen) and fine `dust' particles of heavier elements and molecules left over from earlier cycles of stellar evolution. As the relevant gas cloud collapses, initiated by complex processes of interstellar shock waves, the gas and dust falls into a giant flattened, circulating, pancake-like `protoplanetary' disk. 

Within this disk planets are, we believe, slowly formed by a `bottom-up' process, with bodies of ever-increasing size being produced. Collisions and mergers proceed through a number of stages characterised by qualitative differences in the respective particle interactions. Together these extend over more than 14~orders of magnitude in size, from the original sub-micron dust grains to the giant planets. Specifically, primordial dust particles are believed to progressively collide and stick together, eventually forming rocks of 10~meters or so in size. These then continue to collide and grow over tens of thousands of years to form `planetesimals' (mini-planets) of 10\,km or so in size, and these then grow further due to gravitational attraction, collision, and accretion, through subsequent stages termed runaway, oligarchic, and post-oligarchic growth. The process results in the formation of the rocky `terrestrial' planets (like Earth and Mars), whose internal structures are defined by physical and chemical differentiation, and whose largely spherical shapes are determined as their gravitational forces overcome their material strength. The gas giants (like Jupiter and Saturn) form at larger distances from the host star, as their relatively small terrestrial-type cores rapidly sweep up the residual hydrogen gas left over from the collapsing disk. These gas giants are planets of monstrous size: Jupiter probably comprises a rocky core of just a few Earth masses, overlain by a sea of metallic hydrogen some 70\,000~km deep.

\begin{figure}[t]
\centering
\frame{\includegraphics[width=1.0\linewidth]{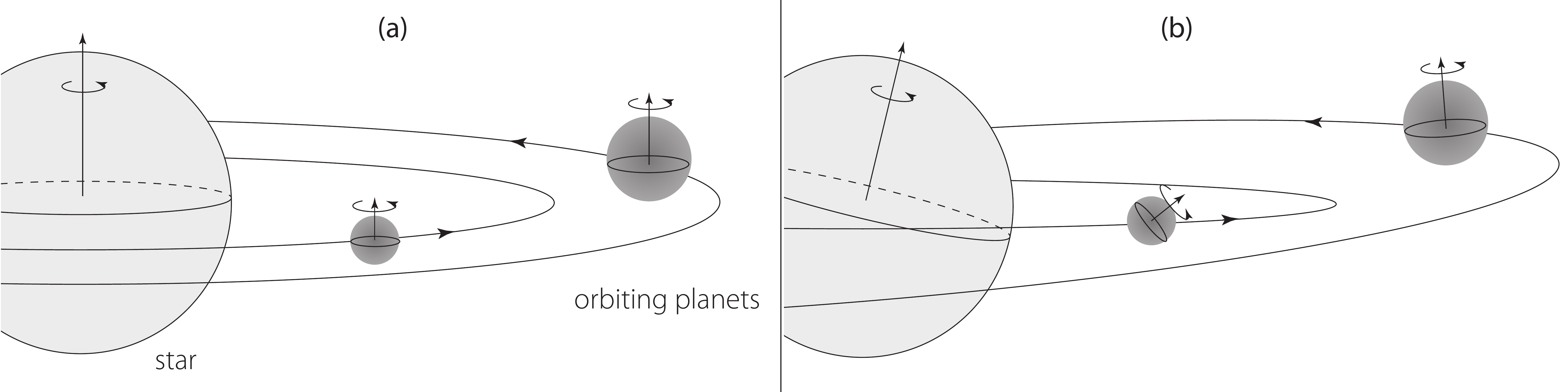}}
\caption{\footnotesize Schematic of different planetary systems. In (a)~the planetary orbits are roughly circular, with their orbital planes coinciding with the host star's equatorial plane, and their spin axes aligned with that of the star (and perpendicular to the orbit planes). In (b)~the two planetary orbits are more eccentric, and neither co-aligned with each other, nor with star's equator. Neither are their spin axes orthogonal to their orbital planes. The solar system shows a general tendency for the types of alignment evident in~(a), although there are a number of misalignments. Much more extreme misalignments are seen in many exoplanet systems.
\label{fig:schematic-solarsystem}
}
\end{figure}

If this entire process proceeds in a fairly orderly manner, then we might expect that the resulting planets will form in circular orbits, with their orbital planes aligned with the rotation axis of the central star, and with their own spin axes perpendicular their orbits. Very broadly, we would expect the more massive planets to form further out from the star where more disk material was available for accretion. Such a process appears likely to have been responsible for the broad features of our solar system (Figure~\ref{fig:schematic-solarsystem}a).

Detailed models of the initial composition of the primordial nebula of gas and dust, combined with thermodynamical and chemical considerations, suggest that molecules would have condensed with distance from the central star according to their degree of volatility: with only the more refractory elements (such as magnesium, silicon, and iron) and molecules (such as magnesium and aluminium silicates) condensing close to the star, with the more volatile elements and molecules only able to condense further out from the star, where the nebula temperatures would be much lower. 

The radius in the protoplanetary disk beyond which water ice can be present is referred to as the `snow line', and it is believed to play an important role in the architecture of planetary systems. Water ice only forms at temperatures below about 180\,K, and the snow line in the early solar nebula probably fell at nearly three times the Sun--Earth distance, consistent with the fact that water-rich (the so-called C-class) asteroids are found predominantly in the outer asteroid belt. Water, it seems, could not have condensed out anywhere close to the Earth's orbit. How then, were the Earth's oceans formed?

\section*{The diversity of exoplanets}
Of more than 500 exoplanets known in early 2011, the majority occur as single planets orbiting a particular star. This does not mean that these stars are only orbited by a single planet, only that for these systems only a single dominant planet, typically of the mass of Jupiter or more, has so far been discovered. The first surprise accompanying the discovery of the first exoplanets was that these early examples of other solar systems were not at all like our own. Very frequently, these planets are massive Jupiter-mass objects, but orbiting seemingly implausibly close to their parent star. Many orbit so close to the star, much closer than Mercury's orbit around the Sun, that their orbital period is only a few days. And while the planets of our solar system orbit our Sun in approximately circular orbits, many of the early exoplanet discoveries had very eccentric orbits.

The first transit measurements of these `hot Jupiters' (so-called because they orbit so close to their parent star that their surface temperatures must be 1000~C or more) showed that their densities are much lower than those of Earth and the other terrestrial planets, and are much more comparable to those of Jupiter and Saturn. Accordingly, they must be composed primarily of hydrogen, the most common element in the Universe, and the predominant element present when the Sun and planets of our own solar system formed some 4.5~billion years ago. Theorists do not believe that these hydrogen-dominated gas giants could possibly have formed in their present orbits so close to the intense heat of their parent stars. Instead, it appears that they must have formed much further out in their own solar system, beyond the orbit of Jupiter in our own. Subsequent to their formation, they must have somehow migrated inwards to their present orbital location. Theorists are presently pre-occupied with understanding when and how the migration occurred, whether the planets have been halted in their inward migration or whether we are seeing a snapshot of inward motions with these mighty giants soon to fall onto their host star, and for how long these giant balls of hydrogen gas can resist the evaporating heat of their parent star.

The orbits of many exoplanets show other bizarre properties. Careful spectroscopic measurements show that their orbital planes can be highly inclined to their host star's equatorial axis, with some planets orbiting in a retrograde sense, that is, with their orbits in the opposite sense to that of their host star's rotational spin. Under certain circumstances, the relative inclination between the orbital planes of two orbiting planes can be measured. While the orbital planes of the solar system planets are rather well aligned, to within typically 2--3~degrees, two of the planets in the Upsilon Andromeda system have a mutual inclination of 30~degrees. These unexpected orbital properties appear not to be extreme cases, but rather common occurrences. The next challenge was to understand how these orbits could have come about, and compelling models of the exoplanet formation and evolution have already been put forward to explain them.

Lower mass and higher density exoplanets are now also being found. By calculating the heat input due to the incident radiation from the host star, we can estimate which of these planets orbit within the so-called `habitable zone', that is, in orbits for which the surface temperature -- dictated by their proximity to their host star -- will allow the presence of liquid water at their surface, being neither too hot nor too cold (the Earth orbits within our solar system's habitable zone: 5\% closer to the Sun and it would experience runaway greenhouse heating, while 5\% further it would experience runaway glaciation). The habitable zone is an important, but just one of, the many special features of our own Earth's characteristics which appears to make it particularly conducive to the long-term development of life: others are its magnetic field which shields high-energy solar radiation, its fortuitous mass, and other features of the Sun itself. Exoplanets which orbit in their own star's habitable zone are starting to be found, and astrobiologists are beginning to develop criteria to assess which of them are likely to be able to support life.

\begin{figure}[t]
\centering
\includegraphics[width=1.0\linewidth]{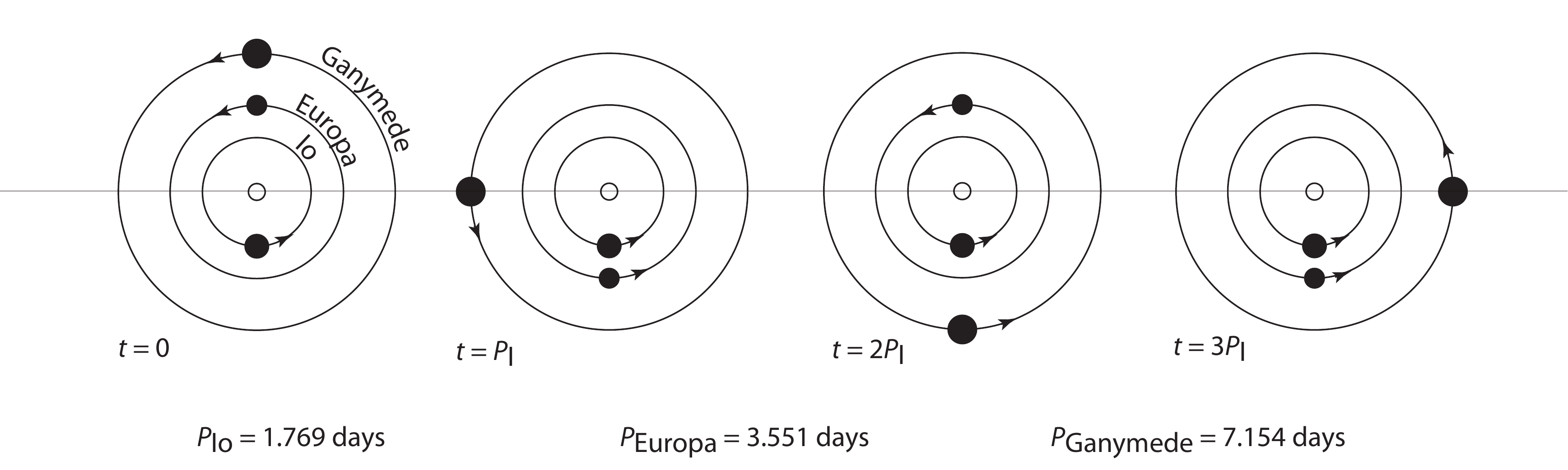}
\caption{\footnotesize The Laplace resonance, in which three orbiting bodies have a simple integer ratio between their orbital periods. Within our solar system, Jupiter's inner satellites Ganymede, Europa, and Io are in such a triple resonance, with Ganymede completing one orbit in the time that Europa makes two, and Io makes four.
\label{fig:laplace-resonance}
}
\end{figure}

\section*{Orbital resonances}
More than 50 relatively nearby stars\endnote{The nearest stars to the Sun are at a distance of around 3--4 light years, meaning that light, travelling at its extravagant 300\,000 kilometers per second, would take 3--4 years to travel from them to us. Stars with planets around them are typically being found at distances of 50--100 light years, although the closest exoplanet system discovered to date lies at just 10~light years. {\it If\/} intelligent life existed there, the round-trip communication time to this system, using light or radio waves, would take some 20~years.} with more than one orbiting planet are now known, and their number is rising rapidly. In late 2010, the star known as HD~10180 was discovered (using the Doppler measurement technique) to be orbited by seven planets, while in early 2011 the Kepler satellite detected a star accompanied by six planets, all of them orbiting very close to their host star, and all of them transiting. Such an arrangement of so many close-in planets came as a total surprise, even to the experts.

Numerous multiple planetary systems show another remarkable property: many of their orbits are in `resonance'. The phenomenon of resonance comprises a strange and complex family of orbital behaviour of celestial objects determined by their mutual gravitational attraction. In the solar system numerous manifestations of resonance are seen. Amongst them are the 3:2 spin--orbit resonance of the planet Mercury discovered by radar observations in 1965, in which Mercury spins about its axis precisely three times in the time taken for it to make two orbits around the Sun. 

`Mean motion resonances' are a particularly important class in which the orbital periods are in a simple integer ratio (such as 2:1, 3:1, 4:1, 3:2, and so on). Examples are the 3:2 Neptune--Pluto resonance (Neptune makes three orbits around the Sun in the 500~years required for Pluto to complete two), the existence of the `Trojan' asteroids preserved by the 1:1 resonance with Jupiter, and the empty lanes in the asteroid belt known as the Kirkwood gaps which correspond to a family of resonances with Jupiter. A specific and most curious case, known already to Laplace, occurs when three bodies orbit in a triple mean motion resonance: Jupiter's inner Galilean satellites Ganymede, Europa, and Io are in a 1:2:4 Laplace resonance, with Ganymede completing 1~orbit in the time that Europa makes~2, and Io makes~4 (Figure~\ref{fig:laplace-resonance}). 

Several of these types of resonance orbits have already been discovered in multiple exoplanet systems. The first triple `Laplace-type' resonance was found, in 2010, for three of the planets orbiting the nearby star known as GJ~876, which have orbital periods of 30.4, 61.1, and 126.6~days. Many more bizarre examples are now predicted to be awaiting discovery. Even retrograde resonances could exist, in which one planet orbits its star in one direction, while another in the same system orbits in the opposite sense. 

It turns out to be reasonably straightforward to understand the physics of how a planetary system will remain locked into a state of orbital resonance once it gets into such a configuration. But it is somewhat less obvious how these states of resonance could have come about in the first place. The explanation lies at the heart of the latest theoretical models which have been developed to explain the formation and evolution of exoplanets, and of our own solar system.

\section*{Disk migration}
Amongst the millions of stars across the heavens, there are those in which star formation and the accompanying protoplanetary disk collapse are seen to be ongoing today. From these, it can be inferred that the process of planet formation -- through the hierarchical assembly of dust, rocks, and planetesimals -- must take place rather rapidly, at least on astronomical timescales. The formation of protoplanets up to 1000~km or so in size is most probably completed within a million years or so. Computer simulations show that, thereafter, there exists an interval of time during which the residual disk of gas out of which the planets formed, and perhaps a residual `sea' of 1000\,km-sized planetesimals, acts as a type of viscous medium which affects the orbits of the planets during the later stages of their formation. Depending on the properties of the protoplanetary disk, the planets can then end up migrating slowly inwards towards their parent star, and sometimes slowly outwards, significant changes to their orbits occurring over several thousand orbital periods. Slightly different speeds of the inward motion of two such migrating planets can then lead to resonant crossings, at which point of time pairs of planets can be locked into a mean motion orbital resonance. Their migrating motion subsequently proceeds in a coupled manner, with the orbital resonance preserved indefinitely thereafter.

During the late stages of the hierarchical assembly of a planetary system, the emerging planets do not exist in isolation, but rather immersed in this cosmic `shooting gallery' of planetesimals and protoplanets competing for their own survival and growth. Over many millions of years, gravitational interactions between the larger bodies can lead to glancing blows and catastrophic impacts which may alter the planet's otherwise stable orbital path. In this chaotic manner, highly elliptical planet orbits, orbits highly inclined with respect to the stellar rotation axis, and planetary spin axes which are significantly shifted with respect to the orbital planes can all result (Figure~\ref{fig:schematic-solarsystem}b). Objects can be pulverised in the process, or they may even escape from the gravitational control of the host star, being ejected from the system through slingshot-type encounters. The final architecture of any given planetary system is likely to be all but unpredictable, and highly sensitive to a plethora of almost arbitrary initial conditions.

\section*{The architecture of the solar system}
Our short tour of the emerging field of exoplanet research, and the theoretical models now in place to explain their formation and evolution, brings us to the point where we can review some aspects of our own solar system, and examine the extent to which these generalised models can help to explain its nature.

Our solar system is a highly complex structure. The Sun, each of the eight planets and their natural moons, and the countless numbers of asteroids and comets, present an overall picture of dazzling complexity. Their mass, density, age, and chemical and isotopic abundances, portray a convoluted history of initial formation and subsequent evolution, whose interactions are controlled by the force of gravity, and sculpted by the phenomenon of resonances. Its understanding has been advanced through meticulous observation and pioneering theories over many centuries, and particularly over the past few decades. The discovery and modeling of exoplanet systems has supplied some additional important clues, and has consolidated a picture of its structure, formation, and evolution. In the remainder of this article, I will consider a few key features of relevance to our own place in the solar system, and set them in the context of our current understanding of its formation.

\vspace{5pt}\noindent
{\it Age of the solar system:} seismological studies of the Sun, and the radiogenic dating of meteorites, place their ages at around 4.57~billion years before present. The Sun, and the first assemblage of bodies formed from the collapsing protoplanetary gas and dust disk, came into being at around this time. 

Meteorites are fragments of (proto)planetary material in the solar system that have crossed the Earth's orbit, and survived their passage through the atmosphere to land on Earth. All known meteorites are fragments of either asteroids, the Moon, or Mars, with the former dominating. They display a large diversity of texture and mineralogy. Most are ancient, dating from the first 10~million years of the solar system formation. They trace key stages of its evolution, such as the composition of the initial interstellar medium, the early stages of the Sun's protostellar collapse, disk formation, dust condensation and coagulation, thermal processing, and planetesimal and planet formation.

The most recent attempts to date the Earth itself, a scientific topic with a long and varied history, fit within this picture. The oldest rocks on Earth, zircon from Jack Hills in Western Australia, are estimated (from their uranium and lead content) to be 4.40~billion years old. Accordingly, the age of the Earth, corresponding to the end of its accretion, its early physical and chemical differentiation, and its core formation, is dated at some 100~million years after the formation of the first meteorites.

\vspace{5pt}\noindent
{\it Formation by hierarchical assembly:} the formation of the eight planets can be well modeled as resulting from the hierarchical assembly of progressively larger and larger bodies -- rocks, planetesimals, and protoplanets -- described already. Closer to the Sun, where the mass of disk material was more limited, and where condensation temperatures would have been high because of the heat from the Sun, the terrestrial planets (Mercury, Venus, Earth and Mars) formed. Further out, with a larger mass of disk material available, and much lower condensation temperatures because of the larger distance from the Sun, terrestrial-type rocky cores formed first, then slowly accumulated their mantle of hydrogen gas and other low-condensation temperature ices.

There are countless numbers of much smaller bodies in the solar system. Planetesimals which grew to modest size without forming larger objects, as well as post-collisional debris, are represented by meteoroids, asteroids, and comets. Meteoroids are mainly `rock' (a combination of iron- and magnesium-bearing silicates and metallic iron) which have irregular orbits and lie between 100~microns and some 10--50\,m in size. Asteroids are the larger rocky bodies which formed inside the orbit of Jupiter. Examples of protoplanets which have survived more-or-less intact are the asteroids Ceres, Pallas and Vesta in the inner solar system, and perhaps the Kuiper belt dwarf planets further out. Comets, often described as `dirty snowballs', comprise frozen ices (water ice, and carbon dioxide), dust grains, and small rocky particles, and range from a few hundred meters to tens of kilometers across.  The `Oort cloud' consists of many billions comets of all orbital orientations, extending out to tens of thousands of times the Earth's orbit, but with a total mass of only a few Earth masses. It originated from the outward gravitational scattering of planetesimals early in the history of the solar system by Uranus and Neptune.

The smallest solar system objects, swarms of sand to small meteoroids, are considered to be impact debris from the final stage of its formation, of which the Moon's impact craters provide evidence. Even now, at intervals of several tens of millions of years, a small planetesimal or comet in an Earth-crossing orbit strikes the Earth. Such impacts may have been responsible for mass extinctions such as at the Cretaceous/Tertiary boundary 65~million years ago.

\vspace{5pt}\noindent
{\it Origin of the Moon:} in the 1890s the astronomer and mathematician George Darwin, the son of the illustrious Charles, hypothesised that the Moon had formed by centrifugal spin-off from the Earth. In the 1970s this was replaced by the idea that the Moon formed as the result of some impact event with Earth. Simulations now provide a plausible dynamical account of such an impact origin, resulting from an oblique and late-stage giant collision between the Earth and a Mars-mass object. These simulations show that the cores of the impacting bodies rapidly coalesced, while the orbital liquid and vapour debris disk solidified and accreted into a single large satellite over an interval of about a thousand years. The idea is further supported by the fact that lunar samples indicate that its surface was once molten. The low iron abundance in the lunar mantle suggests that the impact happened near the end of the Earth's accumulation and differentiation, perhaps a few million years after the solar system formation. The Moon, in short, provides important clues to the turbulent origin of our solar system.

\vspace{5pt}\noindent
{\it Late heavy bombardment:} there is other direct and indirect evidence of the catastrophic impacts that bombarded the bodies of the solar system as it emerged from its early hierarchical formation. The `late heavy bombardment' of the Moon, also known as the `lunar cataclysm', is a period around four billion years ago during which a large number of impact craters are believed to have formed on the Moon, and by inference also on Mercury, Venus, Earth, and Mars. Evidence for the impact clustering includes the radiometric ages of impact melt rocks collected during the Apollo 15--17 lunar missions.

\vspace{5pt}\noindent
{\it Planetary and solar spin axes:} the spins of the solar system planets are neither perfectly aligned nor randomly oriented. Their `obliquities', the angle between their spin and orbital angular momentum vectors (see the schematic in Figure~\ref{fig:schematic-solarsystem}b), range from 0$^\circ$ (Mercury), 3$^\circ$ (Jupiter), 23$^\circ$ (Earth), 25$^\circ$ (Mars), $27^\circ$ (Saturn), $30^\circ$ (Neptune), $98^\circ$ (Uranus), $118^\circ$ (Pluto), and $177^\circ$ (Venus). The original accretion processes, along with more massive late-stage planetesimal or protoplanet collisions, may together explain these large spin axis misalignments -- these huge bodies were, quite simply, knocked about as they formed. The Sun's spin axis is itself misaligned by $7^\circ$ with respect to the mean orbital plane of the planets, and this particular tilt may have originated as a result of a non-uniform infall of material onto the Sun during the later stages of its own formation.

\vspace{5pt}\noindent
{\it The origin of water on Earth:} only beyond the snow line, several times the radius of the Earth's orbit, were the early solar nebula temperatures low enough for water to have condensed. Water in the early solar system was therefore probably initially restricted to its outer regions, i.e.\ to the outer asteroid belt and the region occupied by the giant planets. Recent simulations of the solar system formation suggest that the Earth acquired much of its water oceans, and other surface volatiles, through the various accretion of comets and carbonaceous asteroids throughout its formation history. Asteroids and comets from the Jupiter--Saturn region were probably the first water deliverers, when the Earth was less than half its present mass. The bulk of the water presently on Earth was subsequently carried by a few planetary embryos, originally formed in the outer asteroid belt and accreted by the Earth during the final stages of its formation, perhaps particularly during the `late heavy bombardment' about four billion years ago.  

\vspace{5pt}\noindent
{\it The origin of the Earth's atmosphere:} the Earth's atmosphere comprises volatile gases essential for our very existence which, similarly, would not have condensed {\it in situ}. Three sources have been identified as the possible origin of the atmospheres of the terrestrial planets: the capture of nebular volatile gases, `outgassing' of volatiles locked up in small bodies during collisional accretion, and outgassing from their inner cores and mantles during later tectonic activity (in the case of the Earth, this activity is driven by a combination of its residual heat of formation as well as radioactive decay). The early atmosphere of the Earth most probably originated as a result of accretion, as for its water content. High concentrations of atmospheric carbon dioxide soon after its formation 4.5~billion years ago have progressively decreased with time, firstly some three billion years ago as a result of the appearance of methanogens, methane-producing microbes feeding directly on hydrogen and carbon dioxide, and further around two billion years ago with the appearance of oxygen-producing bacteria. Nevertheless, although evidence for the progressive oxidation of the Earth's atmosphere is well documented geologically, especially since the start of the Phanerozoic eon some 540~million years ago, the details remain uncertain, with volcanic outgassing, and tectonic subduction, all playing a role.

\vspace{5pt}\noindent
{\it Orbital stability and chaos:} measuring the masses, positions, and orbital velocities of the most massive, and therefore the gravitationally most dominant, bodies of our own solar system, allows their positions to be extrapolated to arbitrary times in the past and in the future. However, the problem is far from trivial because the orbits of all objects are continuously being modified, albeit usually only marginally, by close approaches with other objects, which themselves depend on the previous orbits. Very long-term predictions become, in consequence, highly sensitive to even minute effects. Good approximations can be made by numerically calculating the gravitational interactions between all pairs of bodies at any one instant of time, and repeating these calculations over a grid of time steps to estimate their mutual interactions and hence their motions. Limited by computational power, until 1991 the only such numerical integration of a realistic model of the full solar system (the JPL ephemeris DE102), was used to predict the positions of the planets over the (astronomically tiny) interval 1411~BC to 3002~AD.

Developments in computational power, and vastly improved techniques of numerical integration of a large system of self-gravitating bodies, mean that positions of the bodies of the solar system (and in exoplanet systems) can now be carried out billions of years into the past and future. The results provide a wealth of insights into our solar system's past and future architecture, and its delicate dependency on a host of factors. In short, the massive outer planets form a highly stable system, little changing even over these enormous gulfs of time. Adding gravitational `test particles' into these simulations also shows that the solar system is dynamically very `full', in the sense that additional bodies added to the system would escape over much shorter time intervals. 

The inner solar system presents a different picture: small changes in the initial conditions (for example, due to uncertainties on the masses) can lead to orbits in which Mercury will escape from the solar system, or will collide with Venus. Although such extreme events occur with only low probability in these types of simulations, consistent with the continued presence of these bodies in the solar system, such possibilities demonstrate that the solar system is not strictly stable indefinitely. Another general conclusion is that the arrangement of the planets within the solar system, at the end of its formation, was most probably rather different from its present configuration.

\vspace{5pt}\noindent
{\it Planet migration:} understanding the architecture of many exoplanet systems, in particular the existence of the `hot' gas giant Jupiter-type planets orbiting close to their host star, as well as the presence of resonances, suggest that many exoplanets must have migrated, inwards or outwards, after their formation. Furthermore, the fact that gas or planetesimals persist for some time after initial formation, more-or-less demands that such migrations must occur due to the effects of viscous drag. This poses the question: is there evidence for significant planet migration having occurred in the early solar system?

The short answer is `yes'. Such a model postulates a specific starting configuration for the outer solar system planets, which may have existed before an extended phase of planetesimal scattering altered their orbits, and it appears to explain a number of phenomena. In the so-called {\it Nice model}, the giant planets all formed on circular and coplanar (aligned) orbits, Jupiter started marginally further out from the Sun (at 5.45\,AU, compared to 5.2\,AU at present; where 1AU, or `astronomical unit' is the Sun--Earth orbital distance), the other giant planets started in a more compact configuration within about 15--17\,AU, and a sea of planetesimal bodies of total mass 30--50 Earth masses resided interior to 30\,AU with a density falling off with distance from the Sun. The planets subsequently migrated slowly as they moved through the planetesimal sea, exchanging momentum with them. This model can reproduce a number of primary characteristics of the giant planet orbits, namely their present semi-major axes and slight eccentricities, and their mutual inclinations with respect to the mean orbital plane of Jupiter. One of the implications of the increasing eccentricities of the ice giants (Uranus and Neptune) as Saturn crossed the 1:2 Jupiter resonance would have been the penetration of their orbits into the outer planetesimal disk, resulting in a brief phase of rapid and enhanced planetesimal scattering, a cascade of dislodged bodies which may explain the burst of impacts marking the late heavy bombardment of the Moon.

The irregular satellites of the giant planets are dormant comet-like objects that move on both stable prograde and retrograde orbits around them. They may have been dynamically captured during the same planetesimal-dominated orbital migration phase of the outer planets, in which multiple close encounters between the giant planets allowed scattered comets to be captured.

\section*{The future}
Exoplanet research began, in earnest, less than two decades ago. It remains in its infancy, and a golden age of explosive discovery surely lies ahead. Hundreds of exoplanets have already been discovered, and plans are in place for the discovery of perhaps many thousands in the coming decade. These discoveries will probably show that the formation of the solar system was, in one sense, nothing out of the ordinary -- planet formation appears to be a natural by-product of the very process of star formation, which is ubiquitous and ongoing throughout the Universe. Yet, the properties of other planetary systems already appear to be bizarrely diverse. Perhaps the architecture of the solar system will prove to be common and pervasive, or perhaps it will be found to be an almost unique example of a benign arrangement particularly conducive to the development of life. At the present time, we do not know.

The understanding of the origin and the properties of exoplanets has advanced, since their discovery, extremely rapidly, in part because the detailed foundations needed to understand the dynamics and composition of the solar system had already been laid. Nevertheless, their rich diversity has demanded some modifications to the previous theory of the solar system's formation, in particular identifying a phase during which planet migration is a common and inevitable part of the later stages of their hierarchical formation. In turn, planetesimal-driven migration, when applied to the solar system, appears to neatly explain certain details of its current architecture and other characteristics.
 
The broad consistency between the consequences of this migration, and the numerous facets of the complex architecture of the solar system, lends further weight to the overall paradigm of planet formation according to the solar nebula hypothesis, the basics of which were already in place to explain the very existence of our solar system. 

Our present picture, then, is of planets forming in staggering abundance as a more-or-less inevitable by-product of star formation. Microscopic particles of gas and dust, initially orbiting harmlessly in a flattened disk, were swept up to create much more massive bodies that led to a phase of colossal turmoil as the larger planets emerged, migrated, then finally settled down into an arrangement, relatively stable for billions of years, conducive to the development of life.

\begingroup 
\parindent 0pt 
\parskip 3pt 
\def\enotesize{\footnotesize}
\theendnotes   
\endgroup

\section*{About the Author}

Michael Perryman is a Visiting Professor in the Astrophysics Group of the School of Physics, University of Bristol. After a PhD at the Cavendish Laboratory, Cambridge, the major part of his career has been with the European Space Agency where he led the pioneering development of space astrometry  -- the measurement of the positions, distances, and space motions of the stars. He was professor of astronomy in Leiden University between 1993--2009, and spent 2010 as Distinguished Visitor at the University of Heidelberg, and the Max Planck Institute of Astronomy, Heidelberg. He is the recipient of the 1996 Prix Janssen of the Soci\'et\'e Astronomique de France, the 1999 Academic Medal of the Dutch Royal Academy of Arts \& Sciences, an honorary doctorate from Lund University in Sweden in 2010, and the 2011 Tycho Brahe Prize of the European Astronomical Society.

\end{document}